\documentclass{emulateapj}
\usepackage{apjfonts}

\usepackage{natbib}
\usepackage{epsfig}
\usepackage{rotating}

\newcommand{\be}{\begin{equation}}
\newcommand{\ee}{\end{equation}}

\newcommand{\kms}{\mbox{km\,\ensuremath{\rm{s}^{-1}}}}

\shortauthors{Cordiner et al.}

\begin{document}

\title{Mapping the release of volatiles in the inner comae of comets C/2012 F6 (Lemmon) and\\ C/2012 S1 (ISON) using the Atacama Large Millimeter/Submillimeter Array}

\author{M. A. Cordiner\altaffilmark{1,2}, A. J. Remijan\altaffilmark{3}, J. Boissier\altaffilmark{4}, S. N. Milam\altaffilmark{1}, M. J. Mumma\altaffilmark{1}, S. B. Charnley\altaffilmark{1}, L. Paganini\altaffilmark{1,2}, G. Villanueva\altaffilmark{1,2}, D. Bockelee-Morvan\altaffilmark{5}, Y.-J. Kuan\altaffilmark{6,7}, Y.-L. Chuang\altaffilmark{6}, D. C. Lis\altaffilmark{8,9}, N. Biver\altaffilmark{5}, J. Crovisier\altaffilmark{5}, D. Minniti\altaffilmark{10,11}, I. M. Coulson\altaffilmark{12}}


\altaffiltext{1}{Goddard Center for Astrobiology, NASA Goddard Space Flight Center, 8800 Greenbelt Road, Greenbelt, MD 20771, USA.}
\email{martin.cordiner@nasa.gov}
\altaffiltext{2}{Department of Physics, Catholic University of America, Washington, DC 20064, USA.}
\altaffiltext{3}{National Radio Astronomy Observatory, Charlottesville, VA 22903, USA.}
\altaffiltext{4}{IRAM, 300 Rue de la Piscine, 38406 Saint Martin d'Heres, France.}
\altaffiltext{5}{LEISA, Observatoire de Paris, CNRS, UPMC, Universit{\'e}
Paris-Diderot, 5 place Jules Janssen, 92195 Meudon, France.}
\altaffiltext{6}{National Taiwan Normal University, Taipei 116, Taiwan, ROC.}
\altaffiltext{7}{Institute of Astronomy and Astrophysics, Academia Sinica, Taipei 106, Taiwan, ROC.}
\altaffiltext{8}{Sorbonne Universit{\'e}s, Universit{\'e} Pierre et Marie Curie, Paris 6, CNRS, Observatoire de Paris, UMR 8112, LERMA, Paris, France.}
\altaffiltext{9}{Caltech, Pasadena, CA 91125, USA.}
\altaffiltext{10}{Pontifica Universidad Catolica de Chile, Santiago, Chile.}
\altaffiltext{11}{Departamento de Ciencias Fisicas, Universidad Andres Bello, Republica 220, Santiago, Chile.}
\altaffiltext{12}{Joint Astronomy Centre, Hilo, HI 96720, USA.}

\begin{abstract}

Results are presented from the first cometary observations using the Atacama Large Millimeter/Submillimeter Array (ALMA), including measurements of the spatially-resolved distributions of HCN, HNC, H$_2$CO and dust within the comae of two comets: C/2012 F6 (Lemmon) and C/2012 S1 (ISON), observed at heliocentric distances of 1.5~AU and 0.54~AU, respectively. These observations (with angular resolution $\approx0.5''$), reveal an unprecedented level of detail in the distributions of these fundamental cometary molecules, and demonstrate the power of ALMA for quantitative measurements of the distributions of molecules and dust in the inner comae of typical bright comets. In both comets, HCN is found to originate from (or within a few hundred km of) the nucleus, with a spatial distribution largely consistent with spherically-symmetric, uniform outflow. By contrast, the HNC distributions are clumpy and asymmetrical, with peaks at cometocentric radii $\sim500$-1000~km, consistent with release of HNC in collimated outflow(s). Compared to HCN, the H$_2$CO distribution in comet Lemmon is very extended. The interferometric visibility amplitudes are consistent with coma production of H$_2$CO and HNC from unidentified precursor material(s) in both comets.  Adopting a Haser model, the H$_2$CO parent scale-length is found to be a few thousand km in Lemmon and only a few hundred km in ISON, consistent with destruction of the precursor by photolysis or thermal degradation at a rate which scales in proportion to the Solar radiation flux.

\end{abstract}


\section{Introduction}

Astronomical observations of comets provide an important means to study some of the oldest, most pristine material in our Solar System. The bulk of cometary ices are believed to be relatively unprocessed, having accreted at around the time the planets formed (c. 4.5 Gyr ago), and have remained in a frozen, relatively quiescent state since then. Depending on the degree of subsequent thermal and radiative processing, some comets likely contain pristine material from the Solar Nebula or prior interstellar cloud. Studies of cometary ices thus provide unique information on the physics and chemistry of the early stages of the Solar System's evolution. Comets could also have been important for initiating prebiotic chemistry on the early Earth, and their study provides crucial details on the link between interstellar ices and planetary material \citep{ehr00,mum11}. 

Use of gas-phase coma observations as a probe of cometary ice composition requires a complete understanding of the gas-release mechanisms. However, previous observations have been unable to ascertain the precise origins of key coma species including hydrogen cyanide (HCN), hydrogen iso-cyanide (HNC) and formaldehyde (H$_2$CO), and details regarding the possible formation of these species in the coma are not well understood.  Hydrogen cyanide is a trace volatile commonly assumed to originate in the nucleus, but a significant source of HCN in the coma has not yet been ruled out. In comet 103P/Hartley 2, evidence was found for release of HCN from icy grains at nucleocentric distances up to 1000~km \citep{boi14}. Hydrogen iso-cyanide in cometary comae has been discussed as a possible tracer of pristine interstellar material pre-dating the origin of the Solar System \citep{irv96}. However, variations in HNC production rates with heliocentric distance are more consistent with its production in the coma \citep{biv97,irv98,lis08}. Detailed chemical modeling \citep{rod01,rod05} has thus far failed to conclusively identify the origin of cometary HNC. Formaldehyde is ubiquitous in dense interstellar clouds and its study in comets is of major interest to astrochemistry and astrobiology. Production of H$_2$CO in the coma {from a distributed (extended) source} has been observed in several comets \citep{biv99,cot04,mil06}, and a determination of its chemical origin is important for testing the role of comets in delivering prebiotic compounds to the early Earth \citep{oro97,dis06}. 

\begin{table*}
\centering
\caption{Observation parameters \label{tab:obs}}
\begin{tabular}{cccccccccccccc}
\hline\hline
Comet&Setting&Date&UT Time&Int. time$^a$&${r_H}^b$&$\Delta^c$&$\phi^d$&$\bar{\nu}^e$&Ants.$^f$&Baselines$^g$&${\theta_{min}}^h$&PWV$^i$\\
&&&&(min)&(AU)&(AU)&($^{\circ}$)&(GHz)&&(m)&($''$)&(mm)\\
\hline
C/2012 F6&1&2013-06-01&11:52-12:32&29.6&1.47&1.75&35&346.6&30&15.2-1284&$0.88\times0.54$&0.83\\
C/2012 F6&2&2013-06-02&12:11-12:33&17.1&1.48&1.75&35&357.5&28&21.4-2733&$0.78\times0.54$&0.44\\[1mm]
C/2012 S1&1&2013-11-17&11:31-12:15&34.2&0.54&0.88&85&349.8&28&17.3-1284&$0.62\times0.41$&0.57\\
C/2012 S1&2&2013-11-17&12:30-13:27&45.2&0.54&0.88&85&357.2&28&17.3-1284&$0.54\times0.40$&0.52\\
\hline
\end{tabular}
\\
\parbox{0.9\textwidth}{\footnotesize 
\vspace*{1mm}
$^a$ On-source observing time.\\
$^b$, $^c$ Heliocentric distance, Geocentric distance of the comet (JPL Horizons).\\
$^d$ Sun-comet-observer (illumination phase) angle\\
$^e$ Mean observational frequency.\\
$^f$ Number of (12~m) antennae in the telescope array.\\
$^g$ Range of antenna baseline lengths.\\
$^h$ Angular resolution (dimensions of Gaussian fit to PSF) at $\bar{\nu}$, excluding antennae DV07, DV19, DV24 and DV25.\\
$^i$ Median precipitable water vapor column length at zenith.
}
\end{table*}

The most accurate method for determining the production site of a given cometary species is through measurement of its spatial distribution about the nucleus, especially its variation with nucleocentric distance in the innermost coma (within a few thousand km of the nucleus). In this study, we report results from the first cometary observations using ALMA, and present spatially-resolved measurements of the distributions of HNC, HCN and H$_2$CO within the comae of two comets originally from the Oort Cloud reservoir: C/2012 F6 (Lemmon) and C/2012 S1 (ISON).

\section{Observations}
\label{obs}

\begin{table}[t!]
\caption{Detected spectral lines, fluxes, production rates and parent scale lengths  \label{tab:lines}}
\hspace*{-4mm}
\begin{tabular}{llccccl}
\hline\hline
Species&Transition&Frequency&E$_u$&Flux$^a$&Q$^b$&$L_p$$^c$\\
&&(GHz)&(K)&(Jy \kms)&(10$^{26}$\,s$^{-1}$)&(km)\\
\hline
\multicolumn{7}{c}{\bf C/2012 F6 (Lemmon) Setting 1}\\
H$_2$CO     &$5_{1,5}-4_{1,4}$&351.769&62.5&0.86(8)&2.1&$1200^{+1200}_{-400}$\\
HCN         &$4-3$&354.505&42.5&8.55(9)&2.3&$<50$\\[1mm]
\multicolumn{7}{c}{\bf C/2012 F6 (Lemmon) Setting 2}\\
HNC         &$4-3$&362.630&43.5&0.38(9)&$\sim0.1$&\dots$^d$\\[1mm]
\multicolumn{7}{c}{\bf C/2012 S1 (ISON) Setting 1}\\
HCN         &$4-3$&354.505&42.5&11.76(9)&3.5&$<50$\\[1mm]
\multicolumn{7}{c}{\bf C/2012 S1 (ISON) Setting 2}\\
H$_2$CO     &$5_{1,5}-4_{1,4}$&351.769&62.5&3.79(10)&16.4&$280^{+50}_{-50}$\\
HNC         &$4-3$&362.630&43.5&1.92(13)&1.2&$700^{+1100}_{-400}$\\
\hline
\end{tabular}
\\
\parbox{0.9\columnwidth}{\footnotesize 
\vspace*{1mm}
$^a$ Integrated line flux within a 5$''$-diameter circular aperture centered on the comet; $1\sigma$ errors on trailing digits given in parentheses.\\
$^b$ Best-fit production rate from visibility model (errors are $\sim\pm10$\%).\\
$^c$ Best-fit parent scale length, including $\pm1\sigma$ errors. \\
$^d$ Insufficient signal-to-noise for fit.\\
}
\end{table}

Comet C/2012 F6 (Lemmon) is a long-period comet (orbital period approx. 11,000~yr; semi-major axis $a=493$~AU) that reached perihelion on 2013 March 24. Comet C/2012 S1 (ISON) was a sungrazing comet, with $a\gtrsim10,000$~AU, and its orbit passed within 0.013~AU of the Sun at perihelion on 2013 November 28.

Observations were made in Cycle 1 Early Science mode using the ALMA Band 7 receiver, covering frequencies between 338.6 and 364.6~GHz (0.82-0.89~mm). Comet Lemmon was observed 2013 June 1-2 and ISON was observed 2013 November 17 (Table \ref{tab:obs}). The cometary positions were tracked using JPL Horizons ephemerides (JPL\#45 for ISON and JPL\#22 for Lemmon). Two correlator settings permitted simultaneous observation of two sets of spectral lines (plus continuum) for each comet (see Table \ref{tab:lines}). Weather conditions were excellent for all observations, with high atmospheric phase stability, and extremely low precipitable water vapor (zenith PWV~=~0.44-0.83~mm). Quasar observations were used for bandpass and phase calibration. Ceres and Titan were used to calibrate ISON's flux scale, and Pallas was used for Lemmon. The absolute flux calibration error is expected to be less than 15\%. The spatial resolution was 0.4-0.9$''$ (Table \ref{tab:obs}) and the channel spacing was 244~kHz, leading to a (Hanning smoothed) spectral resolution of about 0.42~\kms. 

The data were flagged, calibrated and imaged using standard routines in CASA version 4.1.0 \citep{mcm07}. No signal was detected for baselines $\gtrsim400$~m, so the most distant antennae (DV07, DV19, DV24 and DV25; $>500$~m from the array center) were excluded during imaging. Deconvolution of the point-spread function (PSF) was performed using the H{\"o}gbom algorithm, with natural visibility weighting and a flux threshold of twice the RMS noise in each image. Finally, the deconvolved images were convolved with a Gaussian fit to the PSF.  The continuum peak of comet Lemmon was offset by a (negligible) $0.9''$ NE of image center whereas ISON was offset $6.5''$ NW (explainable as a result of non-gravitational acceleration; \citealt{sek14}). Images of ISON were therefore corrected for the response of the ALMA primary beam (half-power beam-width~$\approx17.5''$). 

Images were transformed from celestial coordinates to cometocentric (projected) spatial distances, the origin of which was determined as the location of peak continuum flux.

\section{Molecular maps and radial profiles}
\label{results}

\begin{figure*}
\centering
\includegraphics[width=0.315\textwidth]{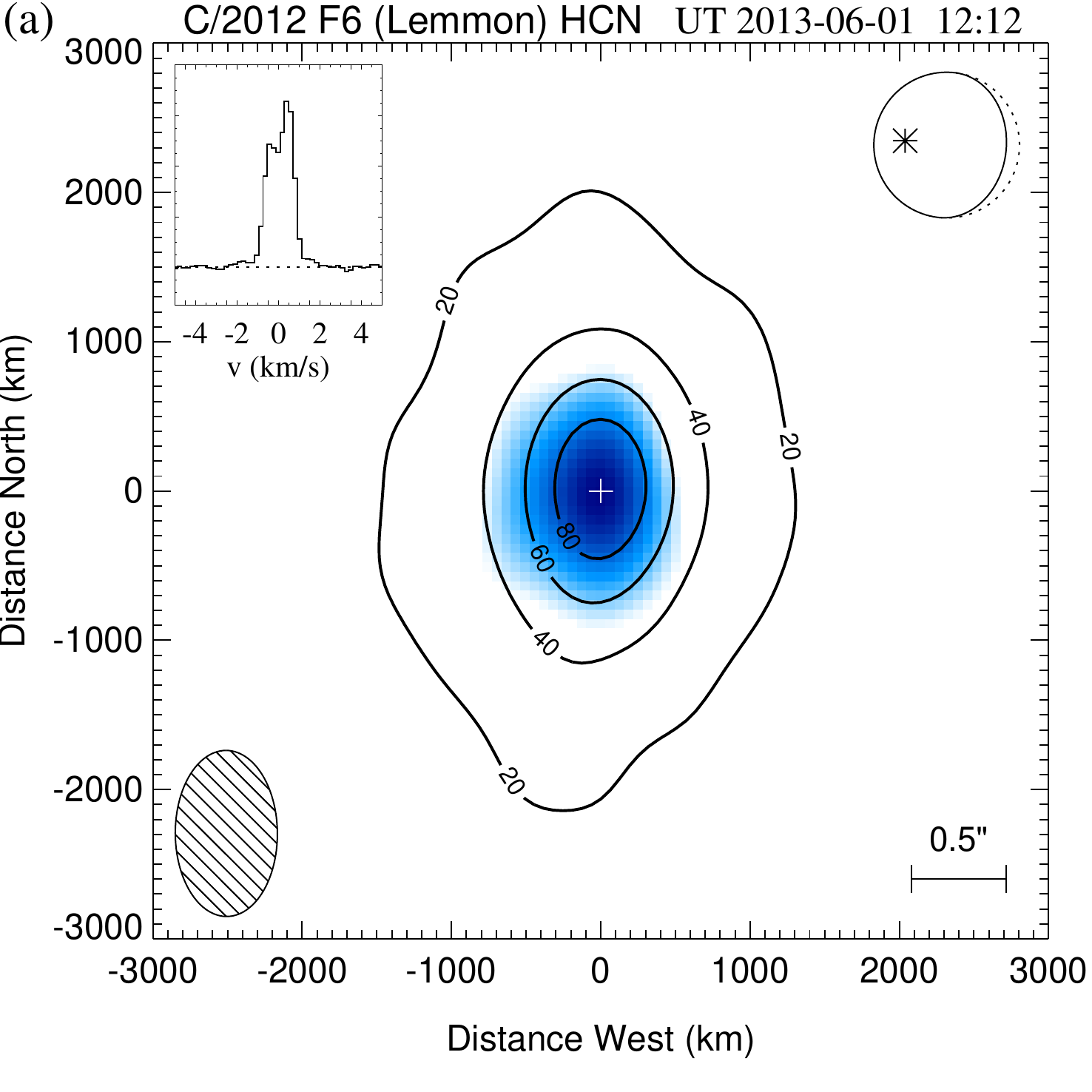}
\includegraphics[width=0.315\textwidth]{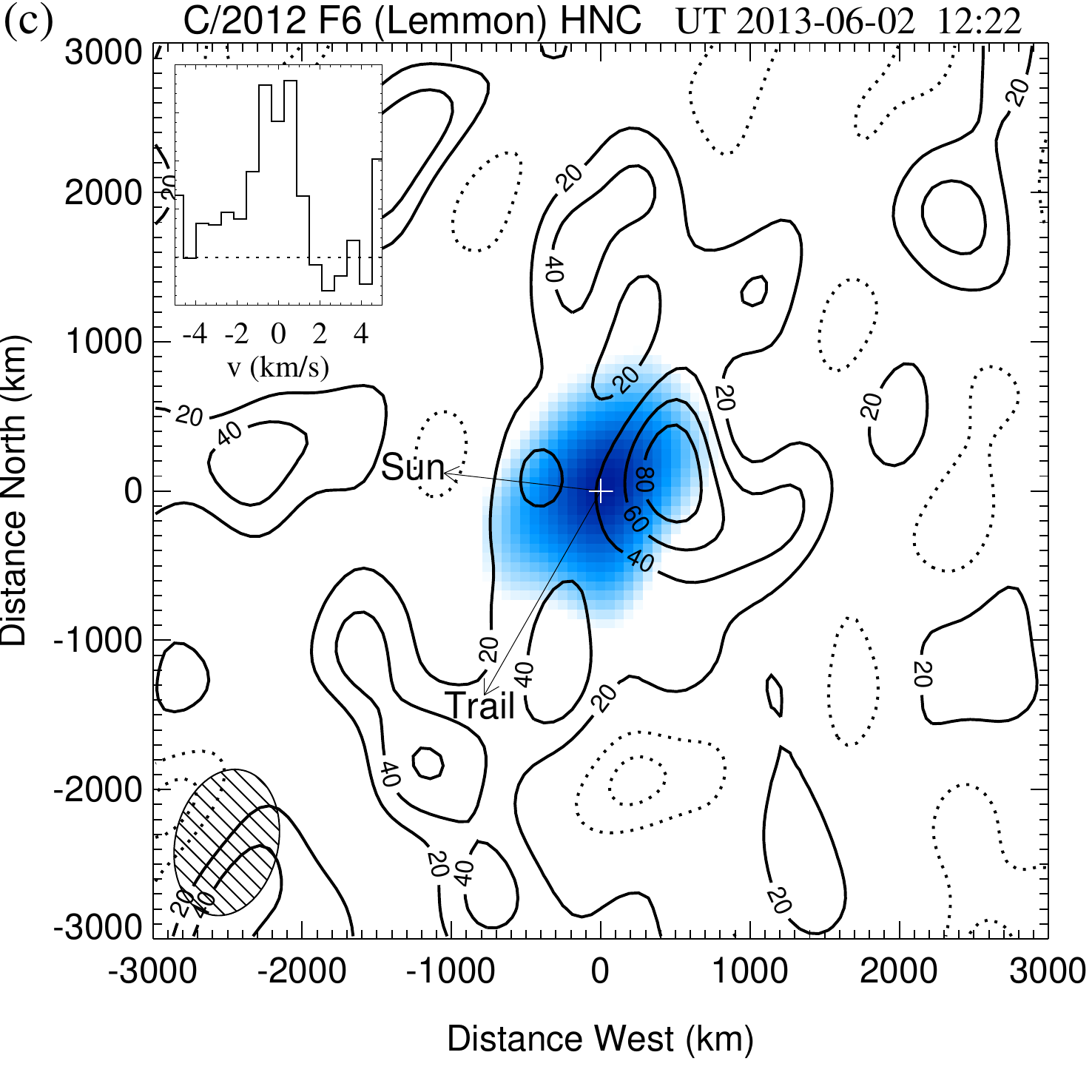}
\includegraphics[width=0.315\textwidth]{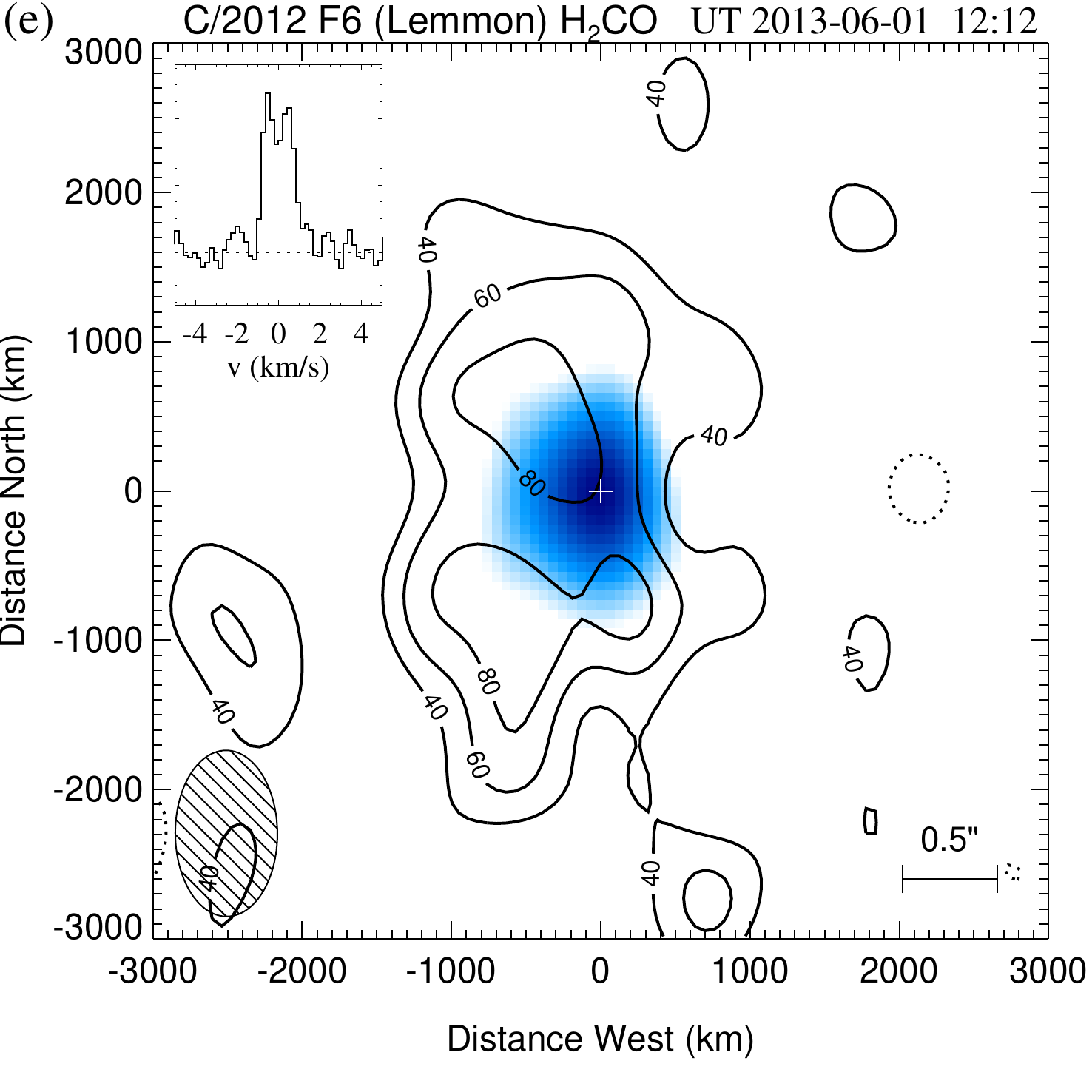}
\includegraphics[width=0.036\textwidth]{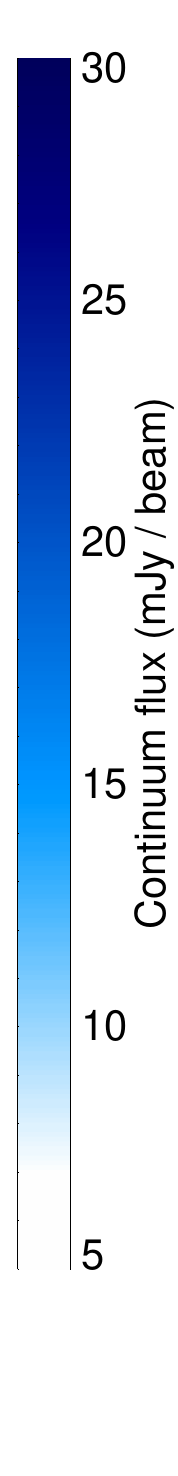}\\
\includegraphics[width=0.315\textwidth]{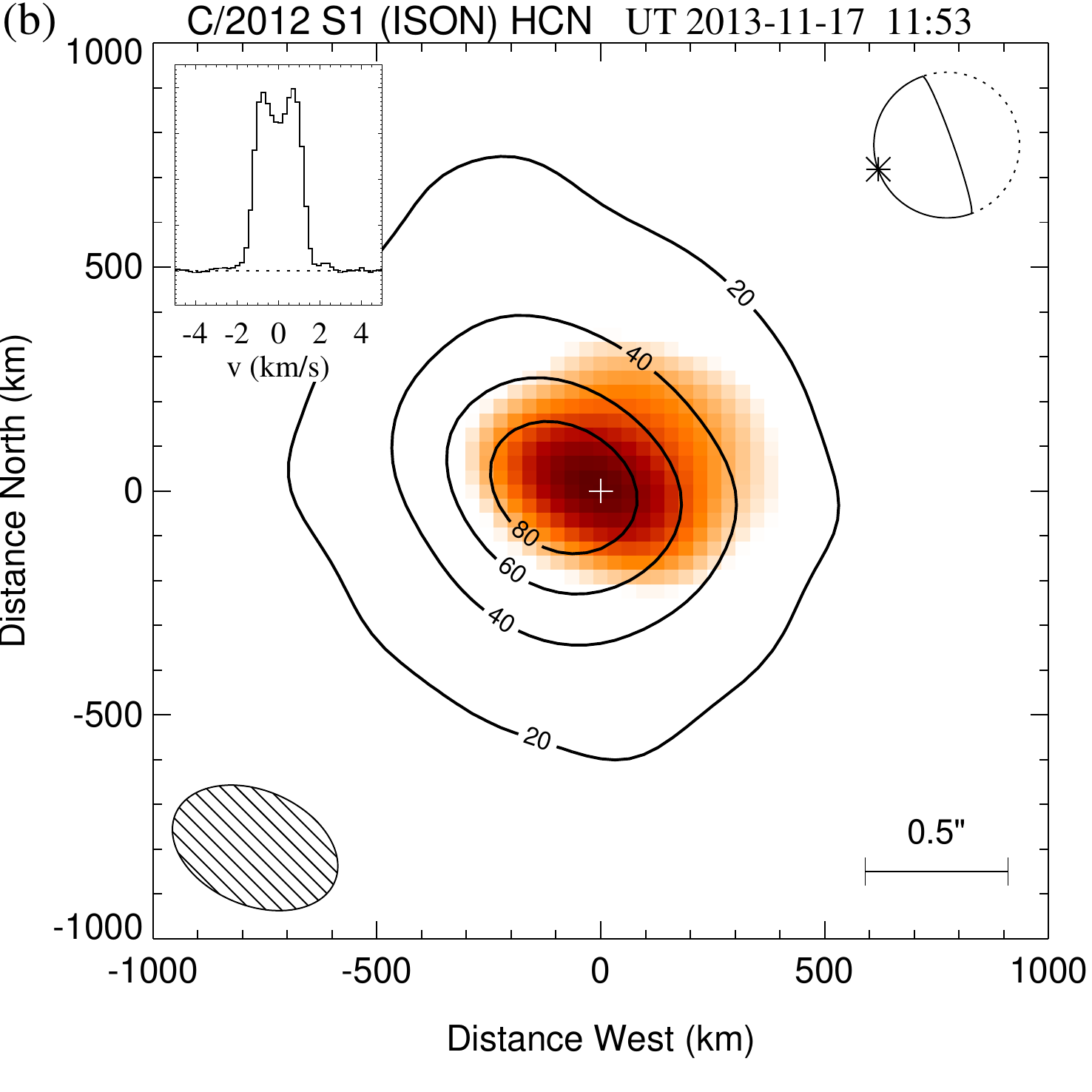}
\includegraphics[width=0.315\textwidth]{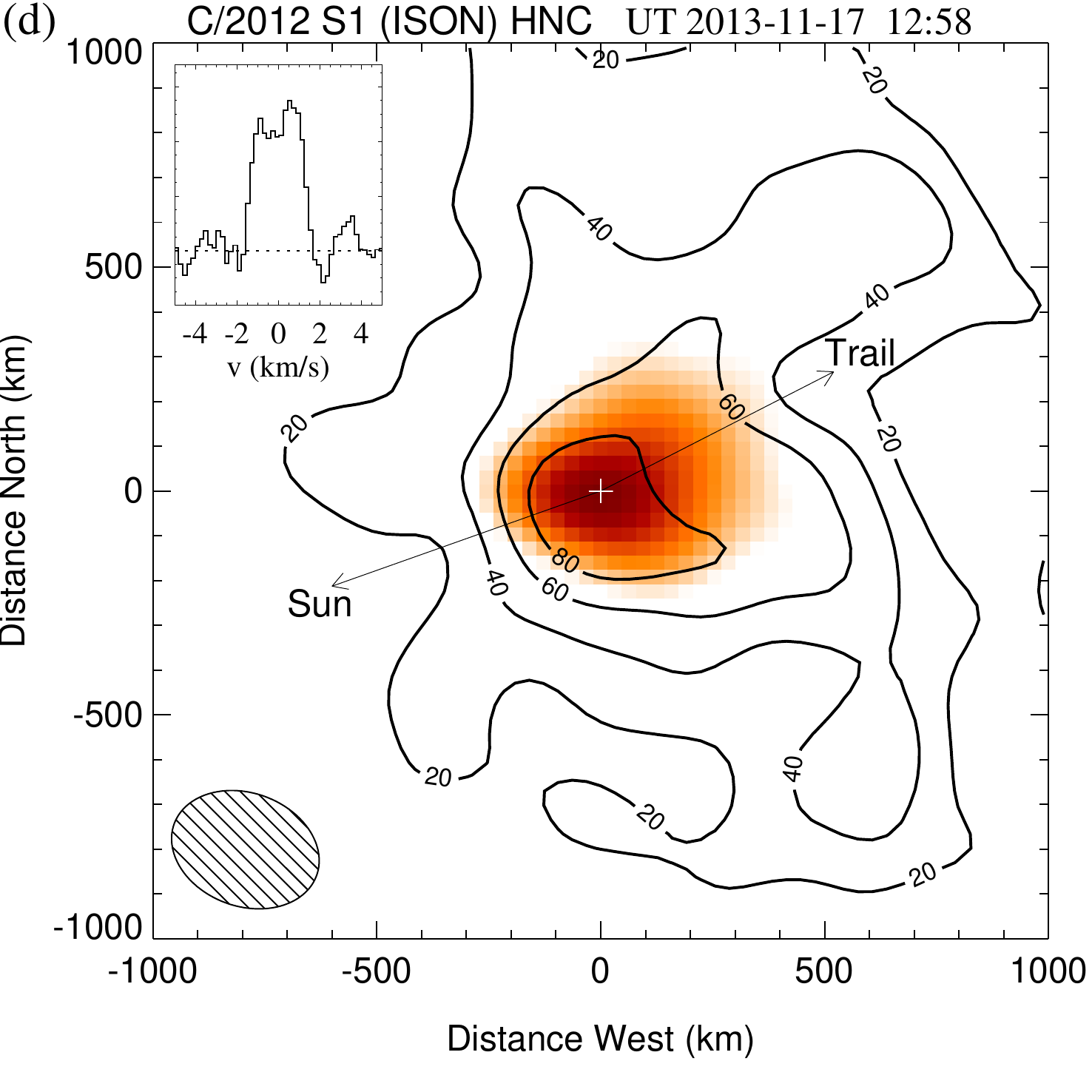}
\includegraphics[width=0.315\textwidth]{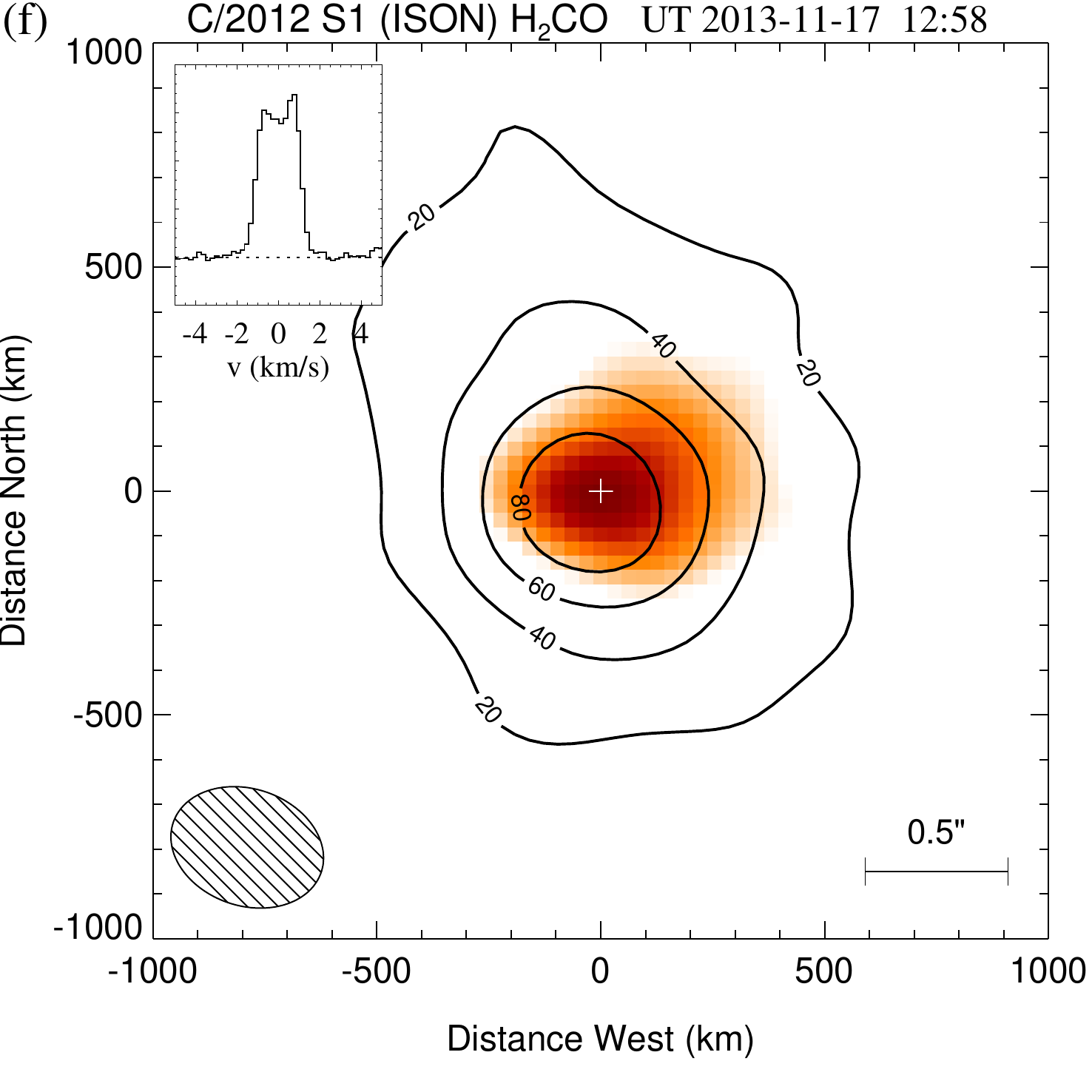}
\includegraphics[width=0.036\textwidth]{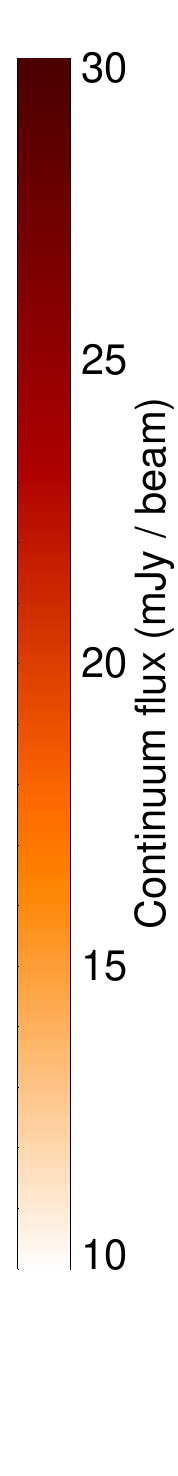}\\
\caption{Contour maps of spectrally-integrated molecular line flux observed in comets F6/Lemmon (top row) and S1/ISON (bottom row). Contour intervals in each map are 20\% of the peak. The 20\% contour has been omitted from panel (e) for clarity. Negative contours are dashed. The RMS noise ($\sigma$, in units of mJy\,beam$^{-1}$\,km\,s$^{-1}$) and contour spacings $\delta$ in each panel are as follows: (a) $\sigma=13.1$, $\delta=19.4\sigma$, (b) $\sigma=10.9$, $\delta=25.2\sigma$, (c) $\sigma=14.8$, $\delta=0.9\sigma$, (d) $\sigma=13.6$, $\delta=2.3\sigma$, (e) $\sigma=13.7$, $\delta=1.1\sigma$, (f) $\sigma=11.0$, $\delta=9.6\sigma$. Simultaneously-observed continuum flux bitmaps are shown in blue for Lemmon and orange for ISON, with flux scales shown far-right. The continuum peaks are indicated with white crosses. Sizes (FWHM) and orientations of the Gaussian point-spread functions are indicated in lower-left (hatched ellipses); observation dates and times are also given. Comet illumination phases ($\phi$) and sub-solar points ($\ast$) are indicated upper-right on panels (a) and (b); projected vectors in the direction of the Sun and the dust trail (the opposite of the comet's velocity vector) are shown on (c) and (d).  Spectra in upper left of each panel show line flux (integrated over the map area) as a function of cometocentric velocity; dashed horizontal lines indicate the flux zero levels.}
\label{fig:maps}
\end{figure*}

The detected spectral lines, including upper-state energies ($E_u$) and integrated fluxes are summarized in Table \ref{tab:lines}. Fig. \ref{fig:maps} shows spectrally-integrated flux contour maps for the observed molecules in each comet, overlaid on bitmap images of the (simultaneously observed) continuum emission.

Dramatic differences are evident between different molecular species observed in the same comet, and between the same species observed in the different comets. By eye, the HCN distributions in both comets appear quite rotationally-symmetric about the central peak. For Lemmon, no offset between the HCN and continuum peaks is distinguishable, whereas ISON's HCN peak is offset 80~km eastward from the continuum peak. Both comets show a compact, strongly-peaked sub-mm continuum, but ISON exhibits an additional, fainter, tail-like feature extending to the north-west, in approximately the opposite direction to the comet's motion (as marked by the `trail' vector in Fig. \ref{fig:maps}d). Cometary sub-mm emission is likely due to large dust grains, $\gtrsim1$~mm in size \citep{jew92}, so ISON's sub-mm tail is consistent with the presence of a debris stream following behind the comet's orbit, as discussed by \citet{sek14}.

For HNC, the emission observed in comet Lemmon (detected at $4.4\sigma$ confidence), is offset from the continuum peak by $500\pm150$~km to the west (Fig. \ref{fig:maps}c) --- in an approximately anti-Sunward direction.  By contrast, ISON's HNC peak (Fig. \ref{fig:maps}d) lies very close to (within 100~km of) the continuum peak. The HNC map for ISON shows a wealth of remarkable extended spatial structure, with at least three streams (identified at $>6\sigma$ confidence), emanating away from the main peak. The majority of HNC emission from both comets is asymmetric, originating predominantly in the anti-sunward hemispheres of their comae.

Formaldehyde also shows strikingly different distributions for comets Lemmon and ISON (Figs. \ref{fig:maps}e and 1f), highlighting the complex nature of this species. Lemmon has a remarkably flat and extended H$_2$CO map, as demonstrated by the size of the region traced by the 40\% contour compared with the other maps. The mean FWHM of the H$_2$CO distribution is $\bar{d}=2920$~km (which is 3.5 times the instrumental PSF value of $\bar{d}=840$~km), and is significantly broader than both HCN and the continuum, which have $\bar{d}=1480$~km and 1180~km, respectively. Lemmon's H$_2$CO map shows two main emission peaks at distances $\sim500$-1000~km from the continuum peak (although noise is likely responsible for some of the structure in this map). By contrast, the H$_2$CO distribution for comet ISON is dominated by a strong, compact central peak (with $\bar{d}=600$~km, compared to $\bar{d}=320$~km for the PSF), and has a relatively symmetrical contour pattern, similar to HCN. 

\begin{figure}
\centering
\includegraphics[width=0.9\columnwidth]{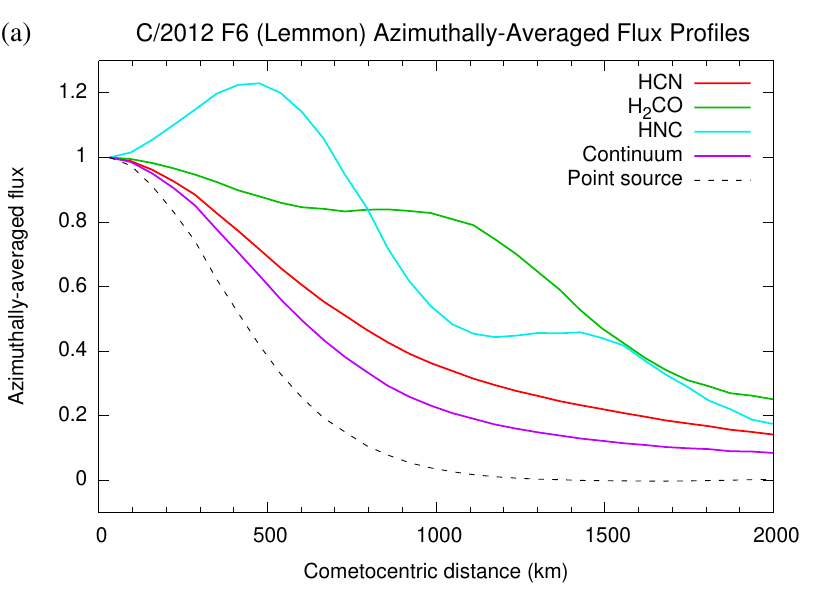}
\includegraphics[width=0.9\columnwidth]{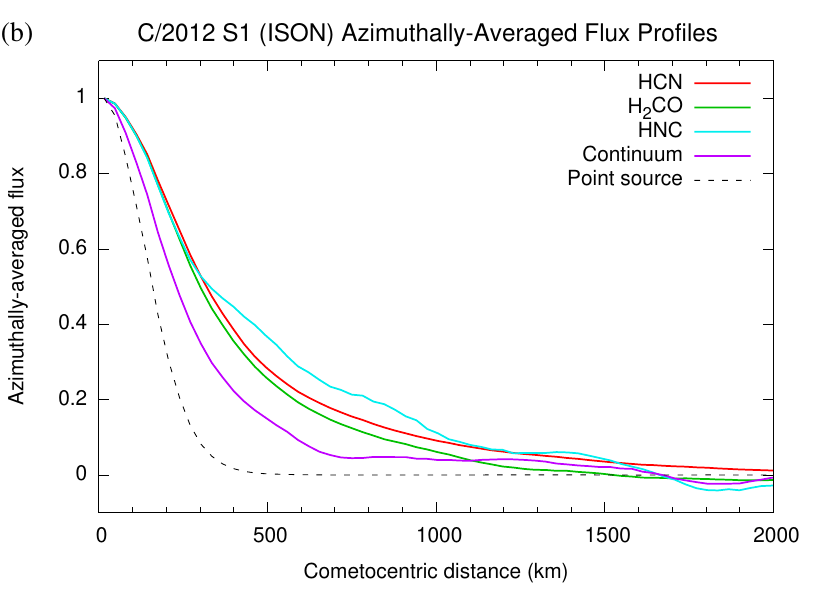}
\caption{Azimuthally-averaged flux profiles for (a) Lemmon and (b) ISON. Profiles have been normalized to unity at the origin. The 354~GHz continuum profile for the (point-source) bandpass calibrators used for each comet are shown with dotted lines.}
\label{fig:profiles}
\end{figure}

Further insight into the flux distributions can be obtained by comparison of the azimuthally-averaged radial flux profiles (Fig. \ref{fig:profiles}). For each map, the HCN flux peak was taken as the origin (apart from the continuum maps, for which the continuum peak was used). The average fluxes inside successive $0.05''$-thick annuli are plotted as a function of annulus radius, normalized to unity in the first annulus.  These azimuthally-averaged profiles are dominated by the $1/\rho$ decay that would be expected from uniform, isotropic expansion (where $\rho$ is the projected radial coordinate). Their shapes are also affected by flux losses due to a lack of short baselines in the array, which become progressively greater for larger structures, and hence, larger cometocentric distances. Despite significant noise ripples, the relatively broad HNC and H$_2$CO profiles for Lemmon (and HNC for ISON) are indicative of coma production for these species. For ISON, the H$_2$CO curve lies inside that of HCN, which is a natural consequence of the shorter lifetime of H$_2$CO (1,280~s \emph{vs.} 23,100~s for HCN at 0.54~AU; \citealt{hue92}). The cause of the relatively narrow continuum profile for both comets will be discussed in more detail in a future article.

\clearpage

\section{Locations of molecular production/release}

\begin{figure*}[t!]
\includegraphics[width=0.33\textwidth]{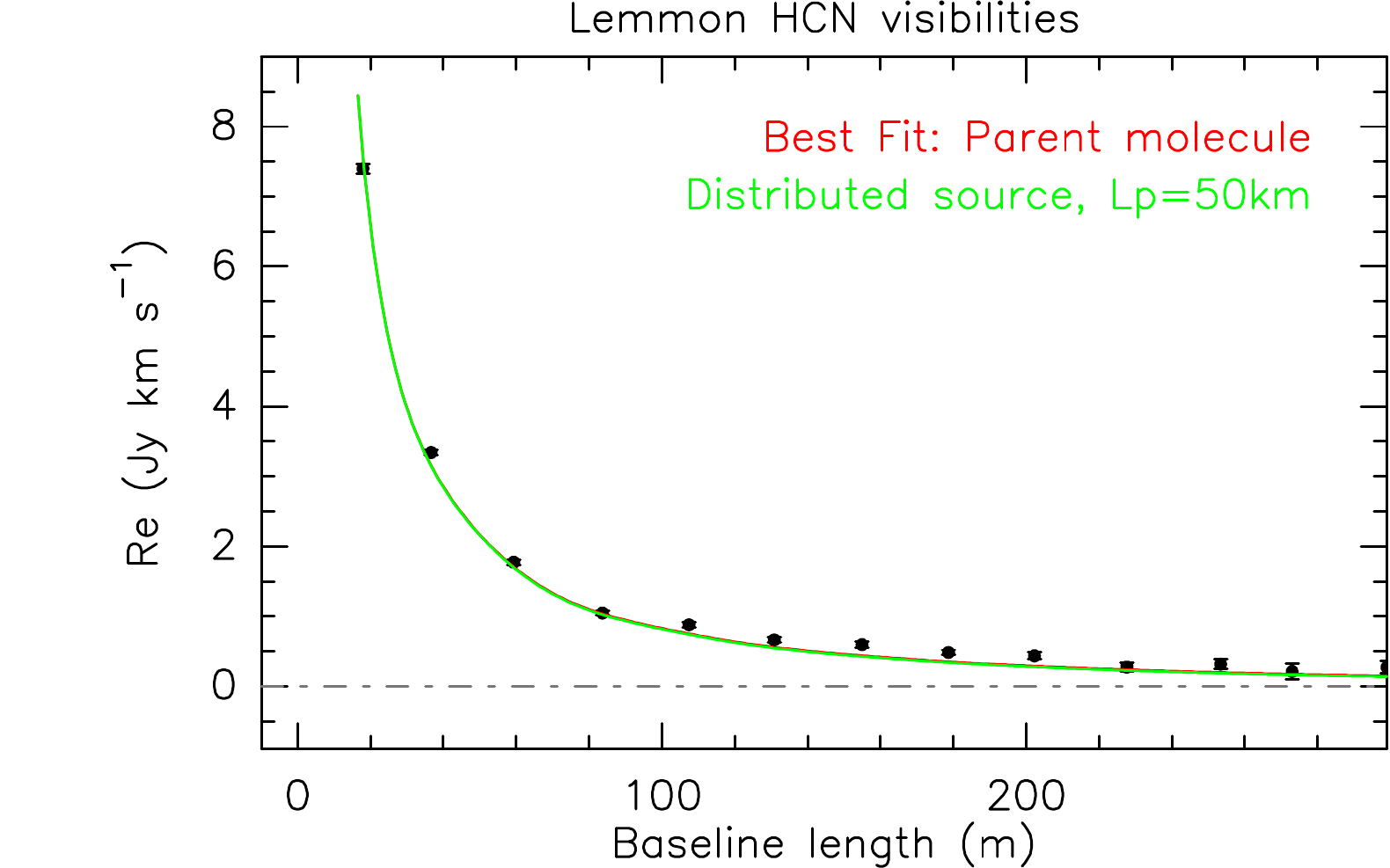}
\includegraphics[width=0.33\textwidth]{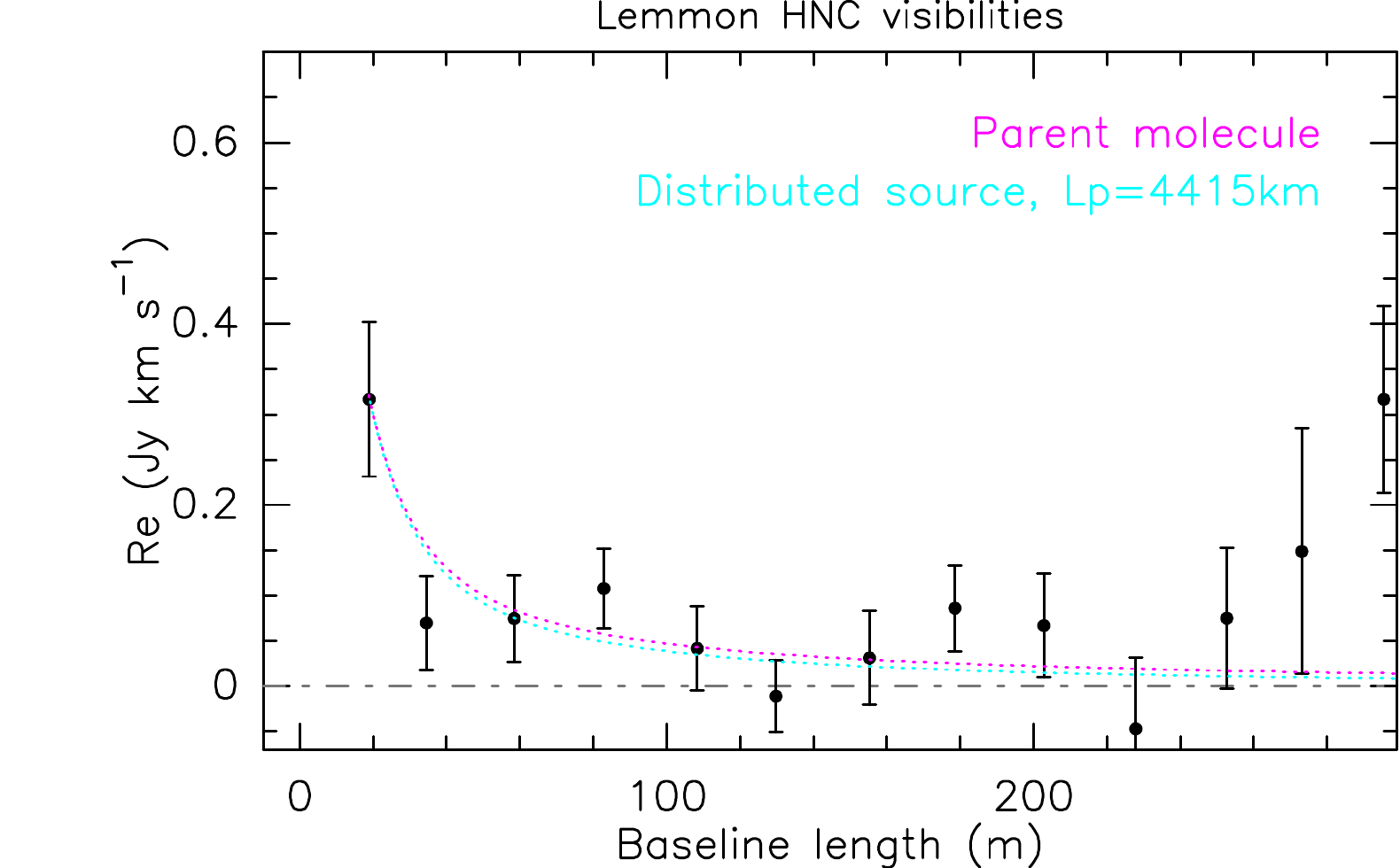}
\includegraphics[width=0.33\textwidth]{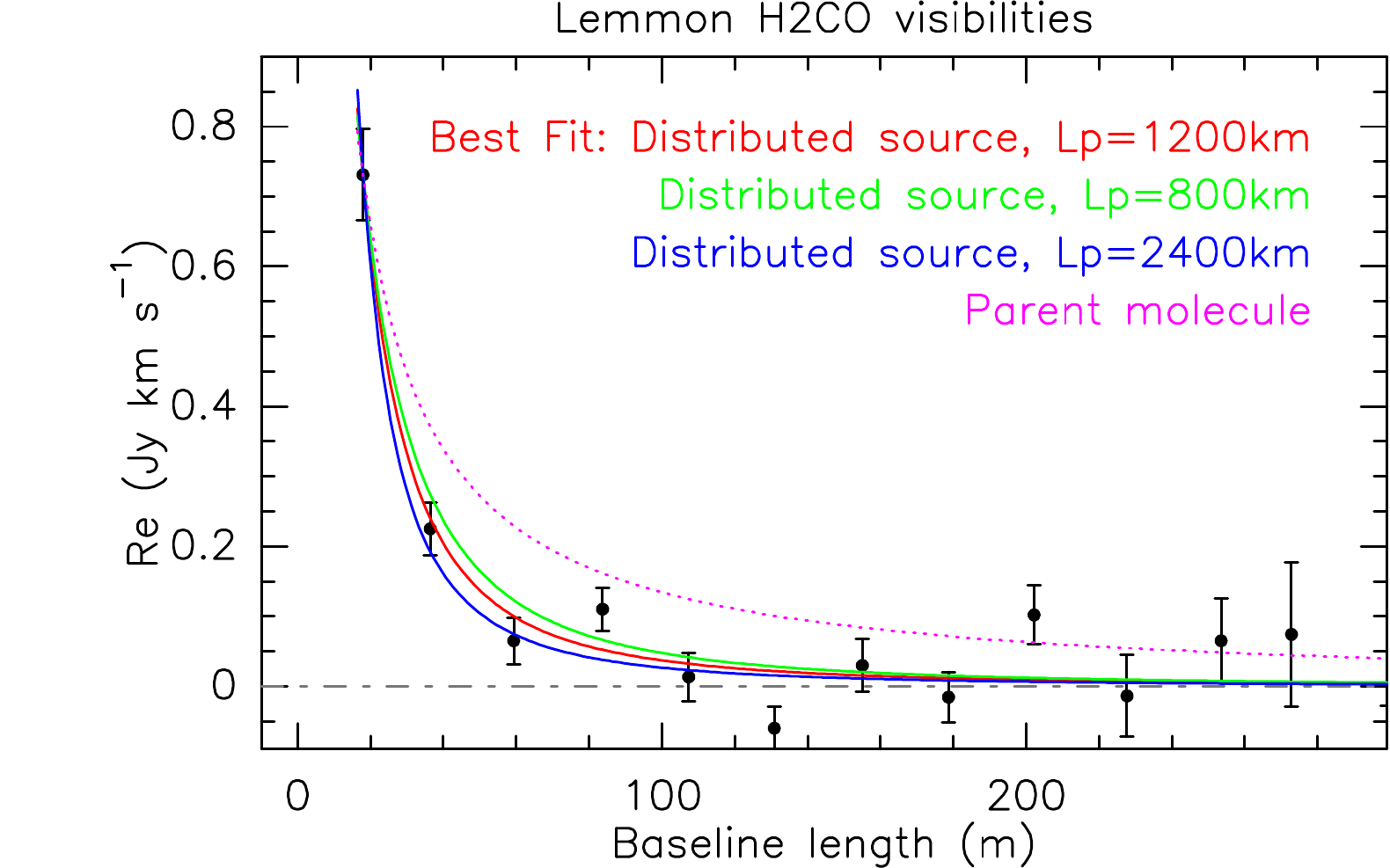}
\\[3mm]
\includegraphics[width=0.33\textwidth]{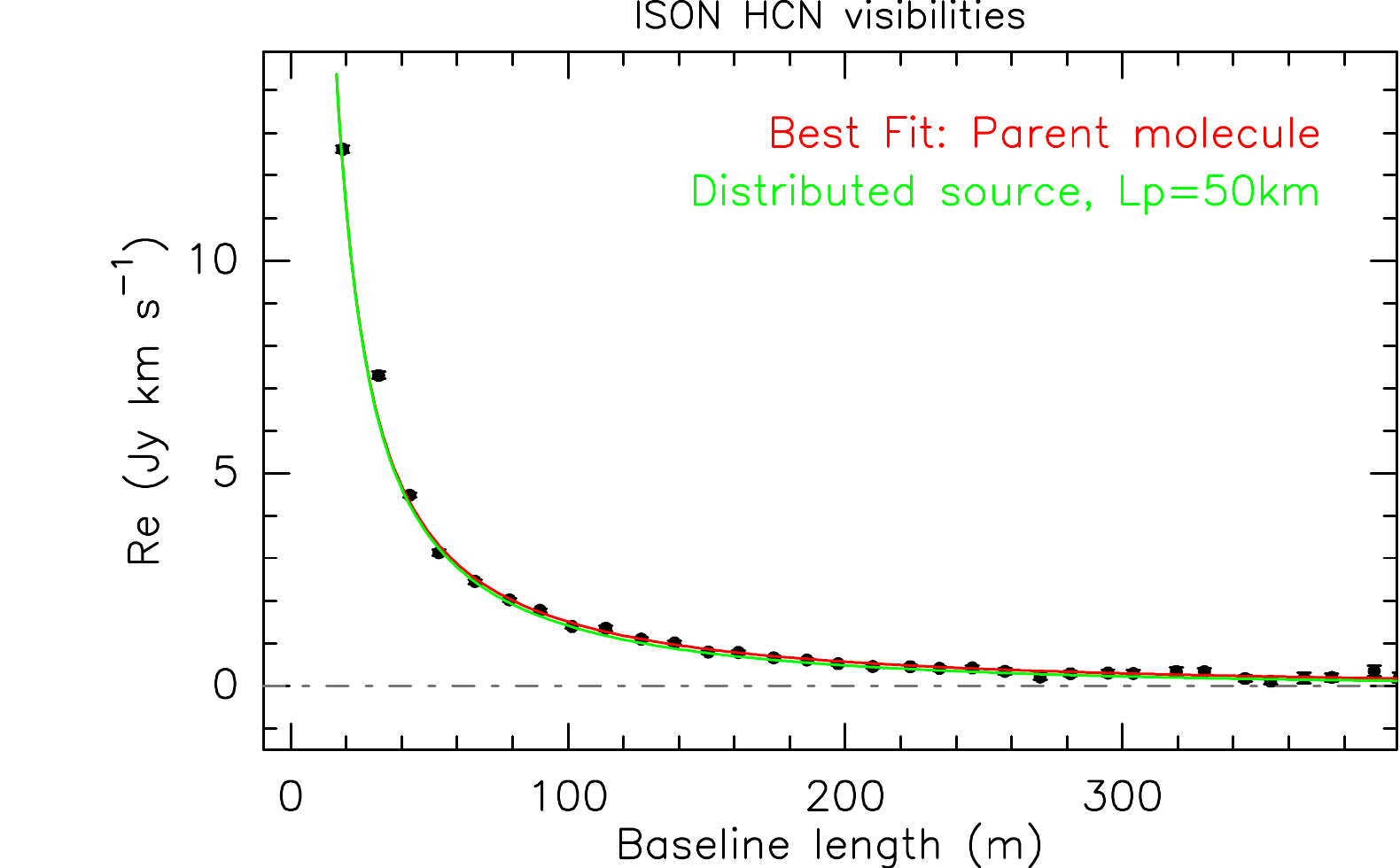}
\includegraphics[width=0.33\textwidth]{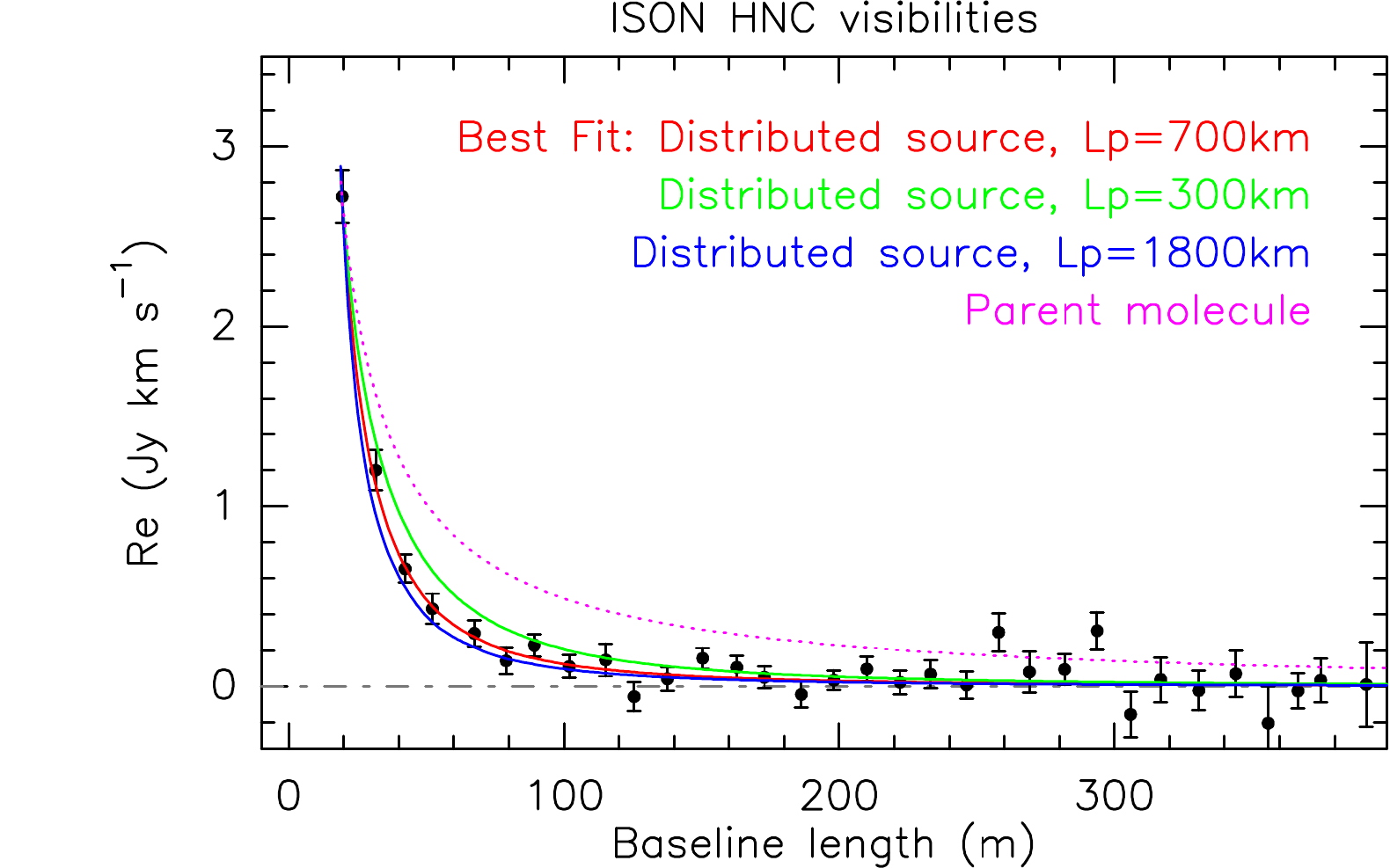}
\includegraphics[width=0.33\textwidth]{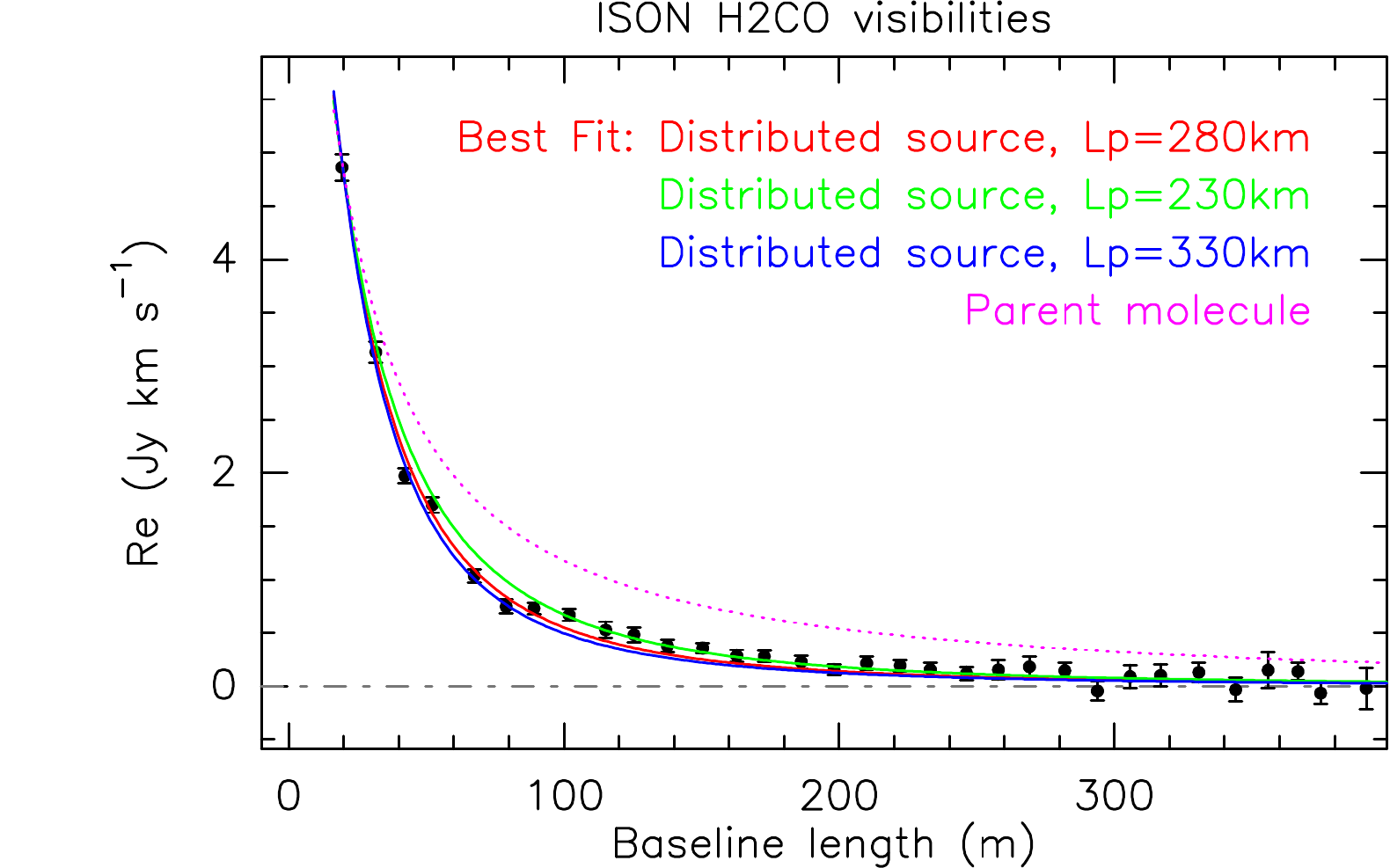}\\
\caption{Real part of observed visibility amplitude ($Re$) \emph{vs.} baseline length for each molecule (phase center set to HCN peak position). Model visibility curves are overlaid, calculated for various parent scale-lengths ($L_p$). The best-fitting curves are shown in red; $\pm1\sigma$ error margins on $L_p$ are shown in blue and green, respectively; purple dotted style indicates best-fitting parent ($L_p=0$) curves in those cases for which a {distributed} source ($L_p\neq0$) fits best.}
\label{fig:vis}
\end{figure*}

Following the method of \citet{boi07,boi14}, the interferometric visibility amplitudes for the observed molecules were modeled under the assumption of uniform, isotropic outflow. In this \citep{has57} paradigm, each molecular distribution is determined by the production rate ($Q$), outflow velocity ($v$), parent scale length ($L_p$) and (for daughter species) the photodissociation rate, with $Q$ and $L_p$ as free parameters.  Respective outflow velocities of $1.0$~\kms\ and $0.7$~\kms\ for ISON and Lemmon were obtained from the HWHM of the HCN lines. For ISON, a gas kinetic temperature $T=90$~K was adopted \citep{agu14}. For Lemmon, $T=55$~K was obtained by scaling the measurement of \citet{biv13} assuming $T\propto r_H^{-1}$. The molecular excitation calculation considers collisions with H$_2$O and electrons, and pumping by Solar infrared radiation. Goodness of fit was determined by minimizing the sum of the squares of the differences between the real parts of the observed and modeled visibility amplitudes. The best-fitting models for the observed visibilities are shown in Fig. \ref{fig:vis}, and corresponding $Q$ and $L_p$ values are given in Table \ref{tab:lines}.

\subsection{HCN and HNC}

The presented maps and visibilities provide clear evidence regarding the origins of the observed molecules. The comparison between HCN and HNC is particularly revealing. As shown in Fig. \ref{fig:vis}, our visibility data are most consistent with production of HCN as a primary species, released from (or very near to) the nuclei of both comets. By contrast, for HNC in comet ISON a {distributed} source is required (with $L_p=300$-1800 at $r_H=0.54$~AU).  The HNC visibilities in comet Lemmon are also consistent with a {distributed} source, although the error bars are large, so a parent model fits this data equally well.

Fig. \ref{fig:maps}d shows significant quantities of HNC in streams/clumps at projected distances up to 1000~km from the nucleus of comet ISON that are not present in HCN. These suggest that HNC is released in collimated/anisotropic outflows. The presence of offset HNC emission in Lemmon's coma is also consistent with this hypothesis. From interferometric observations of comet C/1995 O1 (Hale-Bopp), \citet{win97} found evidence for coma production of HNC, and \citet{bla99} identified HNC release in jets, whereas HCN was released from (near to) the nucleus. Variations in excitation cannot plausibly explain the observed differences between these molecules because of the near-identical upper-state energies of the observed HCN and HNC transitions (Table \ref{tab:lines}). 

Our measured HCN production rate of $2.3\times10^{26}$~s$^{-1}$ in comet Lemmon (at $r_H=1.47$~AU on 2013 June 1) is in agreement with that observed by \citet{pag14} (at $r_H=1.74$~AU on 2013 June 20). For ISON, our value of $3.5\times10^{26}$~s$^{-1}$ on 2013 November 17 is consistent with the value of $3.6\times10^{26}$~s$^{-1}$ measured by \citet{agu14} two days earlier. Our $Q({\rm HNC})/Q({\rm HCN})$ ratio of 34\% is somewhat larger than the value of 18\% found by \citet{agu14}, which may be indicative of variations in the relative HCN and HNC production rates with time.  The $Q({\rm HNC})/Q({\rm HCN})$ ratio in comet Lemmon was only about 0.4\%, which is a factor of $\sim85$ less than in comet ISON. In a sample of 14 moderately active comets, \citet{lis08} found a similarly strong change in this ratio with heliocentric distance.  The strong dependence of HNC production rate on $r_H$ \citep[see also][]{irv98}, and its asymmetric spatial distribution, imply release of HNC from a refractory component of the nucleus, ejected into the coma in anisotropic streams.

\subsection{H$_2$CO}

Formaldehyde (H$_2$CO) has a clear {distributed} source in both comets. Visibility modeling reveals a parent scale-length of $L_p=800$-2400~km for Lemmon and $L_p=230$-330~km for ISON (corresponding to photodissociation rates of $\Gamma\approx6\times10^{-4}$~s$^{-1}$ and $\approx4\times10^{-3}$~s$^{-1}$, respectively). Considering the eight-fold increase in Solar insolation expected at ISON's heliocentric distance ($0.54$~AU) compared with Lemmon's ($1.47$~AU), the order-of-magnitude difference in $\Gamma$ is consistent with optically-thin photodissociation of the parent in a uniform outflow. This suggests that the parent of H$_2$CO could be an (unknown) organic molecule or simple-addition polymer of low number \citep{cot08}.

Several previous studies have identified H$_2$CO release in the coma \citep[\emph{e.g.}][]{biv99,cot04}, and our result is qualitatively similar to those. The parent scale lengths we derived are smaller than previous estimates, which were in the range (4000-8000)$r_H^{1.5}$ km. The results may, however, be consistent given that our observations probe only the inner few thousand km of the coma, whereas previous studies probed distances $\gtrsim2000$~km. Due to the lack of short baselines, our ALMA data cannot rule out the presence of additional H$_2$CO sources with angular sizes $\gtrsim5''$. We also note that, for previous radio measurements, observational and model uncertainties (particularly the H$_2$CO excitation), were significant; the latter are less important for our observations, which probe the densest part of the coma where departures from local thermodynamic equilibrium are minimized.

The H$_2$CO production rates we derived for Lemmon (Table \ref{tab:lines}) are compatible with the upper limit of $2.5\times10^{26}$~s$^{-1}$ from \citet{pag14} on June 20. ISON's H$_2$CO production rate was 6.9 times greater, despite only a factor of 1.5 increase in HCN. Similar to HNC, previous observations \citep{fra06} identified a stronger $r_H$ dependence for $Q({\rm H_2CO})$ than for $Q({\rm HCN})$, which is consistent with our results, although the difference in $Q({\rm H_2CO})$ between Lemmon and ISON could also be attributed (at least in part) to a difference in the chemical composition of the nucleus.

By considering the thermal degradation properties of polyoxymethylene (POM) embedded in organic grains, {\citet{cot04} and \citet{fra06}, respectively, successfully modeled the observed H$_2$CO parent scale length in 1P/Halley and the H$_2$CO production rate as a function of $r_H$ in Hale-Bopp}. The scale-length of POM depends strongly on the sizes and temperatures of the grains. Using the model of \citet{fra06} assuming a grain outflow velocity and temperature that vary as $r_H^{-0.5}$, for Lemmon, our observed H$_2$CO scale length is consistent with grains a few microns in diameter, whereas for ISON, grain sizes $\gtrsim40$~${\mu}m$ are required. In reality, an ensemble of grain sizes, temperatures and velocities will be present, necessitating detailed modeling to confirm if thermal degradation of POM matches our H$_2$CO observations.

\section{Conclusion}

We have presented ALMA measurements of molecular line and continuum emission from two Oort Cloud comets: C/2012 F6 (Lemmon) and C/2012 S1 (ISON). These data reveal the detailed spatial structures and origins of HCN, HNC, H$_2$CO and dust within the innermost few thousand km of the cometary comae. For both comets, the dominant source of HCN was from (or very near to) the nucleus. By contrast, the HNC distributions suggest production from a source entering the coma in anisotropic, clumpy stream(s). A {distributed} H$_2$CO source was identified in both comets. The scale-length of the putative H$_2$CO parent was on the order of a few hundred km for comet ISON and a few thousand km for Lemmon, consistent with a parent destruction rate that scales with the intensity of Solar radiation. Relative to HCN, comet ISON's coma was about an order of magnitude richer in H$_2$CO and HNC than comet Lemmon, consistent with a more rapid production of these molecules at ISON's smaller heliocentric distance.

The release of HNC and H$_2$CO as product species implies the existence of chemical precursor materials in the coma, which undergo sublimation, photochemical and/or thermal degradation to produce the observed molecules in the gas phase.  Heating or photolysis of refractory materials such as dust grains, polymers or other macro-molecules, and their subsequent breakdown at distances $\sim100$-10,000~km from the nucleus presents the most compelling hypothesis for the origin of the observed H$_2$CO and HNC. The presence (and composition) of the hypothesized macro-molecular precursors will be measured by the COSIMA instrument on the Rosetta spacecraft during its encounter with comet 67P/Churyumov-Gerasimenko in 2014 \citep{kis07,ler12}. 

The interferometric data presented in this Letter show that routine high-resolution observations of the distributions of molecules and dust grains in cometary comae are now possible. These observations pave the way for future measurements of spatially, spectrally and temporally-resolved coma emission, from which presently little-understood properties such as the detailed physical structure of the coma (on size scales of several hundred km), and the molecular release and reaction mechanisms, will be derived.

\acknowledgments
This work was supported by the NASA Astrobiology Institute through the Goddard Center for Astrobiology, and NASA's Planetary Atmospheres and Planetary Astronomy Programs. It makes use of the following ALMA data: ADS/JAO.ALMA \#2012.A.00020.S and \#2012.A.00033.S. ALMA is a partnership of ESO (representing its member states), NSF (USA) and NINS (Japan), together with NRC (Canada) and NSC and ASIAA (Taiwan), in cooperation with the Republic of Chile. The Joint ALMA Observatory is operated by ESO, AUI/NRAO and NAOJ. The National Radio Astronomy Observatory is a facility of the National Science Foundation operated under cooperative agreement by Associated Universities, Inc. DCL is supported by NASA through JPL/Caltech. DM is supported by Basal CATA PFB-06 and the ICM MAS. YJK is supported by NSC grants 99-2112-M-003-003-MY3 and 100-2119-M-003-001-MY3.

\end{document}